%% file: main.tex
\documentclass[10pt,conference]{IEEEtran}
\IEEEoverridecommandlockouts
\usepackage{amsmath,amssymb}
\usepackage{cite}
\usepackage{booktabs}
\usepackage{array}
\usepackage{float}
\usepackage{algorithm}
\usepackage{algpseudocode}
\usepackage{xcolor}
\usepackage{tikz}
\usepackage{pgfplots}
\usepgfplotslibrary{groupplots}
\usetikzlibrary{positioning,arrows.meta,calc,plotmarks}
\input{figures/plot_style}

\usepackage{eso-pic}

\newcommand{\TopOp}[1]{\mathop{\mathrm{Top}\text{-}#1}\displaylimits}

\title{Index-Free Dynamic Edge Retrieval with Energy-Tail-Aware Partial Scans}

\author{\IEEEauthorblockN{Mohammad Arif Rasyidi}
\IEEEauthorblockA{\textit{Department of Computer Science} \\
\textit{Khalifa University}\\
Abu Dhabi, United Arab Emirates \\
100066916@ku.ac.ae}
\and
\IEEEauthorblockN{Omar Alhussein}
\IEEEauthorblockA{\textit{Department of Computer Science} \\
\textit{Khalifa University}\\
Abu Dhabi, United Arab Emirates \\
omar.alhussein@ku.ac.ae}}

\begin{document}
\maketitle
\bstctlcite{IEEEtran:BSTcontrol}

\AddToShipoutPictureFG*{%
  \AtPageLowerLeft{%
    \raisebox{0.20in}{%
      \hspace*{0.65in}%
      \makebox[0pt][l]{%
        \footnotesize\copyright\ 2026 IEEE%
      }%
    }%
  }%
}

\begin{abstract}

	Dynamic maximum inner-product search (MIPS) returns the $K$ stored vectors with
	the largest dot products with a query while allowing the dataset to change
	through insertions, replacements, and deletions. For edge retrieval, the
	challenge is to achieve high recall and fast queries without making updates
	expensive. Full-vector scanning keeps updates simple but compares each query
	with every stored vector, while indexed methods reduce query cost at the
	expense of maintaining additional structures during updates. We propose ETAR,
	an index-free method that reduces query work while preserving simple updates.
	ETAR keeps the query coordinates with the largest squared values until they
	cover most of its total squared magnitude and treats the rest as a
	low-magnitude tail. It estimates similarity from the retained coordinates using
	a compact lower-precision representation, corrects for skipped coordinates, and
	reranks a fixed number of candidates using full-precision vectors. Across five
	runs on nine static datasets, ETAR averages 99.2\% Recall@10, the fraction of
	exact top-10 results recovered, while running over 4$\times$ faster than exact
	scanning at a representative setting. This speedup also extends to an ARM-based
	mobile device, where ETAR is up to 6.9$\times$ faster across four synthetic
	distributions. Under five streaming workloads, it maintains 100\% Recall@10 at
	every measured point without index rebuilds. Overall, ETAR offers a practical
	middle ground for dynamic MIPS by reducing query cost while retaining simple,
	index-free updates. Code is available at https://github.com/arasyi/etar-mips.

\end{abstract}


\input{sections/01_intro}
\input{sections/02_system_model}
\input{sections/02_method}
\input{sections/03_experiments}
\input{sections/04_results}
\input{sections/05_limitations}
\input{sections/06_conclusion}

\bibliographystyle{IEEEtran}
\bibliography{references}
\end{document}

%% file: figures/plot_style.tex
\definecolor{etarBlue}{HTML}{0072B2}
\definecolor{etarTeal}{HTML}{009E73}
\definecolor{etarOrange}{HTML}{E69F00}
\definecolor{etarPurple}{HTML}{CC79A7}
\definecolor{etarSky}{HTML}{D55E00}
\definecolor{etarGold}{HTML}{B58A00}
\definecolor{etarGray}{HTML}{555555}
\definecolor{etarBlack}{HTML}{111111}

\tikzset{
	exact mark/.style={draw=etarBlack,mark=*,mark options={solid,draw=etarBlack,fill=etarBlack},mark size=2.5pt},
	full8 mark/.style={draw=etarOrange,mark=square*,mark options={solid,draw=etarOrange,fill=etarOrange},mark size=2.3pt},
	etar mark/.style={draw=etarBlue,mark=triangle*,mark options={solid,draw=etarBlue,fill=etarBlue},mark size=2.8pt},
	etarlm mark/.style={draw=etarTeal,mark=diamond*,mark options={solid,draw=etarTeal,fill=etarTeal},mark size=2.6pt},
	hnsw mark/.style={draw=etarPurple,mark=triangle*,mark options={solid,rotate=180,draw=etarPurple,fill=white},mark size=2.8pt},
	ivf mark/.style={draw=etarSky,mark=square,mark options={solid,draw=etarSky,fill=white},mark size=2.4pt},
}

\pgfplotsset{
	compat=1.18,
	paper axis/.style={
			axis line style={draw=black!72,line width=0.45pt},
			tick style={draw=black!72,line width=0.45pt},
			tick align=outside,
			major grid style={draw=black!12,line width=0.30pt},
			grid=major,
			label style={font=\footnotesize},
			tick label style={font=\scriptsize},
			legend style={
					font=\scriptsize,
					draw=black!24,
					fill=white,
					fill opacity=0.96,
					text opacity=1,
					inner xsep=3pt,
					inner ysep=2pt,
					row sep=-1pt,
					cells={anchor=west}
				},
			legend cell align=left,
			scaled ticks=false,
			clip marker paths=true,
			every axis plot/.append style={line width=0.85pt},
		},
	exact line/.style={/tikz/exact mark},
	full8 line/.style={/tikz/full8 mark,dash pattern=on 3pt off 1.4pt},
	etar line/.style={/tikz/etar mark},
	etarlm line/.style={/tikz/etarlm mark,densely dashed},
	hnsw line/.style={/tikz/hnsw mark,densely dotted},
	ivf line/.style={/tikz/ivf mark,dashdotted},
	annoy mark/.style={only marks,draw=etarGold,mark=x,mark options={solid,draw=etarGold},mark size=2.7pt,line width=0.9pt},
	scann mark/.style={only marks,draw=etarGray,mark=star,mark options={solid,draw=etarGray,fill=etarGray},mark size=2.6pt},
}

%% file: sections/01_intro.tex
\section{Introduction}

Applications on edge devices use local vector retrieval for personalization,
recommendation, sensor matching, retrieval-augmented inference, and on-device
memory. Compared with large server-side systems, edge devices often face
tighter compute, memory, bandwidth, and maintenance constraints. Their datasets
can also change over time as new observations arrive, stale items are removed,
or user-specific content is replaced. We focus on dynamic maximum inner-product
search (MIPS), which returns the indices of the $K$ stored vectors with the
largest dot-product similarity to a query as the dataset changes through
inserts, replacements, and deletes. MIPS is a common objective in vector
retrieval and related approximate search
systems~\cite{ramMaximumInnerproductSearch2012,shrivastavaAsymmetricLSHALSH2014a,neyshaburSymmetricAsymmetricLSHs2014,phamSimpleEfficientAlgorithms2021}.

A simple baseline is a direct full-vector scan that computes a complete dot
product between the query and every stored vector and returns the rows with the
largest scores. This approach is attractive for dynamic datasets because
updates modify the stored vectors directly, with no auxiliary search index to
maintain. The drawback is query cost. Each query performs coordinate-wise
arithmetic over the whole dataset and accesses the corresponding vector values,
so cost grows with both dataset size and vector dimension.

Approximate nearest-neighbor (ANN) methods reduce query cost by avoiding
exhaustive comparisons with every stored vector. Hashing methods use randomized
or data-dependent hash functions, including symmetric and asymmetric
constructions for MIPS, to retrieve likely neighbors
efficiently~\cite{indykApproximateNearestNeighbors1998,charikarSimilarityEstimationTechniques2002a,andoniOptimalDataDependentHashing2015,shrivastavaAsymmetricLSHALSH2014a,neyshaburSymmetricAsymmetricLSHs2014}.
Tree and graph indexes guide search through organized neighborhoods or
partitions of the vector
space~\cite{bentleyMultidimensionalBinarySearch1975,mujaScalableNearestNeighbor2014,malkovEfficientRobustApproximate2020}.
Inverted-file and quantized retrieval systems reduce search cost through coarse
partitioning, product quantization, or low-precision distance
computation~\cite{babenkoInvertedMultiIndex2015,jegouProductQuantizationNearest2011,douzePolysemousCodes2016,andreQuickerADCUnlocking2021,guoAcceleratingLargeScaleInference2019,gaoRaBitQQuantizingHighDimensional2024,johnsonBillionScaleSimilaritySearch2021}.
These methods are effective for static or mostly static workloads, but their
auxiliary structures must be built, tuned, stored, and maintained as the data
change, increasing maintenance cost.

We propose Energy-Tail-Aware Reranking (ETAR), an index-free method for dynamic
MIPS that retains the update simplicity of full-vector scanning while reducing
query work. ETAR uses a two-stage pipeline: candidate generation forms a
shortlist likely to contain the top-$K$ vectors, and reranking computes exact
dot products for those candidates. For each query, ETAR treats each
coordinate's squared magnitude as its energy, retains the coordinates with the
largest squared values until they cover a target fraction of the query's total
energy, and treats the rest as a low-energy tail. During candidate generation,
it computes partial scores on the retained coordinates using a compact
low-precision representation and applies a tail-aware correction for the
skipped coordinates. These scores are used only for candidate selection, while
final ranking uses full-precision stored vectors. Although ETAR still considers
all stored vectors during candidate generation, it reduces per-vector work and
avoids auxiliary index rebuilds. Its coordinate selection is related to
adaptive and sampling-based MIPS
methods~\cite{tiwariFasterMaximumInner2022,bruchApproximateAlgorithmMaximum2024},
while the overall design combines query-energy selection, tail-aware scoring,
fixed-budget reranking, and update-friendly storage.

Our contributions are:
\begin{itemize}
	\item We introduce ETAR, an index-free dynamic MIPS method that combines
	      query-dependent coordinate selection, tail-aware candidate scoring, and
	      fixed-budget reranking to reduce query work while preserving simple updates.

	\item We evaluate ETAR across nine static datasets, five streaming workloads, an
	      ARM-based mobile device, and scales of up to one million stored vectors,
	      measuring retrieval quality, query latency, update cost, maintenance overhead,
	      and memory footprint against exact scanning and indexed ANN baselines.
\end{itemize}

%% file: sections/02_system_model.tex
\section{System Model and Problem Description}
\label{sec:problem}

We consider a dynamic retrieval system that stores vectors in a local table and
processes similarity queries as the table changes over time. At time $t$, the
table has $N_t$ occupied rows among $C_t\geq N_t$ allocated slots. Each
occupied row $i=1,\ldots,N_t$ stores a vector $x_i\in\mathbb{R}^d$ and its
metadata. A row is active if it has not been marked as deleted. Let \(
A_t\subseteq\{1,\ldots,N_t\} \) denote the set of active rows when a query is
issued. We omit the subscript $t$ when the time is clear from context, and
assume that queries are issued only when $A_t$ is nonempty.

For scores $s_i$ over a finite set $B$, we use \( \TopOp{L}_{i\in B} s_i \) to
denote the indices of the $\min\{L,|B|\}$ largest scores, with ties resolved by
a fixed deterministic rule. The operator returns row indices rather than score
values.

Given a query $q\in\mathbb{R}^d$, the score of active row $i$ is $s_i(q)=q^\top
	x_i$. Dynamic maximum inner-product search (MIPS) then retrieves the active-row
indices with the largest scores, denoted by
\begin{equation}
	I_K(q;A_t)
	=
	\TopOp{K}_{i\in A_t} q^\top x_i.
	\label{eq:mips}
\end{equation}
Here, $I_K(q;A_t)\subseteq A_t$ is the exact target index set. The query
pipeline assumes $q\neq 0$. If $q=0$, all rows have the same score, so the
fixed tie rule determines the exact target.

The table evolves through inserts, replacements, and deletes. An insert appends
a new row and marks it active. A delete marks an active row inactive without
immediately removing its storage. A replacement deletes the old row and appends
the new vector as a fresh active row. Thus, stale or deleted rows may remain
stored until compaction, but they must not be returned by a query.

The objective is to maintain the vector table under this update stream while
answering queries over the current active set $A_t$. ETAR aims to return an
approximate index set $\widehat{I}_K(q;A_t)\subseteq A_t$ that closely matches
the exact target in \eqref{eq:mips}, while keeping ordinary updates local and
reducing query, update, and maintenance costs.

%% file: sections/02_method.tex
\section{ETAR Design}
\label{sec:method}

ETAR is index-free: it does not build a graph, tree, or inverted file. Its
candidate-generation pass visits table rows directly but scores them using only
query-selected coordinates in an 8-bit view. The main ETAR configuration then
computes full dot products for a fixed number of candidates using 32-bit stored
vectors. The low-memory variant, ETAR-LM, uses row-major 8-bit codes for the
final reranking stage.

\subsection{Representation and Dynamic Tables}

ETAR stores two views of each row. The compact scan view is used during all-row
candidate generation and stores signed 8-bit integer codes in column-major
order, together with one scale per row, the row norm, a deletion marker, and a
row identifier. The main ETAR configuration also stores the 32-bit
floating-point vector in row-major order for exact reranking. ETAR-LM replaces
this full-precision view with a row-major copy of the 8-bit codes. Candidate
generation is unchanged, but the final ordering is approximate.

For a stored vector $x_i$, ETAR defines the row scale and signed 8-bit codes as
\begin{equation}
	\begin{aligned}
		\gamma_i & =
		\begin{cases}
			\max_j |x_{ij}|/127, & \|x_i\|_2>0, \\
			1,                   & \|x_i\|_2=0,
		\end{cases}    \\
		c_{ij}   & =\operatorname{clip}\!\left(
		\operatorname{round}(x_{ij}/\gamma_i),-127,127\right).
	\end{aligned}
	\label{eq:quantization}
\end{equation}
Row-wise scaling keeps updates simple: an inserted or replacement vector can be
encoded independently, without retraining a global quantization codebook.

The compact 8-bit codes are stored in column-major order, so all rows for a
coordinate are contiguous in memory. This layout matches ETAR's query-dependent
coordinate selection: after the selected coordinates are known, ETAR streams
only those coordinate columns across rows. The reranking view is row-major
because reranking accesses all coordinates of only the shortlisted rows.

Most row updates are local. Inserts append the scan view, reranking view, and
metadata. Deletes set the deletion marker, so deleted rows remain stored until
compaction but are ignored during candidate selection and final output.
Replacements mark the old row deleted and append the new vector as a fresh
active row. Tail scoring statistics are updated with the normalized squared
coordinates of the inserted or deleted row, giving $O(d)$ ordinary-update cost.
A reference dynamic-table policy doubles capacity $C$ on overflow and compacts
active rows into capacity $C/2$ when $|A|\leq C/4$. Under this policy, the
$O(Cd)$ copies are amortized over intervening updates. Ordinary updates do not
rebuild a search index.

\subsection{Query Pipeline}

ETAR uses two query stages. Candidate generation forms a shortlist using the
compact scan view, and reranking scores a fixed number of shortlisted rows
using the row-major view. The reranking budget satisfies $R\geq K$.

For a nonzero query $q$, let $H_h(q)$ contain the $h$ coordinates with the
largest squared query values $q_j^2$. The retained query energy is
\begin{equation}
	m_h(q)=\frac{\sum_{j\in H_h(q)} q_j^2}{\|q\|_2^2}.
	\label{eq:retained-energy}
\end{equation}
Given a retained-energy target $\rho$ and a cap $h_{\max}$, ETAR chooses the
smallest $h$ such that $m_h(q)\geq\rho$, capped at $h_{\max}$. If the target is
not reached before the cap, ETAR sets $h=h_{\max}$. We denote the selected
coordinates by $H$ and the skipped coordinates by $\bar H$.

Candidate generation first computes the partial score on the selected
coordinates as
\begin{equation}
	\widehat{p}_i(q)=\gamma_i\sum_{j\in H} q_j c_{ij}.
	\label{eq:reduced-score}
\end{equation}
The same query values are used for coordinate selection, tail scoring, and
exact reranking.

The partial score in \eqref{eq:reduced-score} ignores skipped coordinates,
which can still affect the true dot product. ETAR therefore defines the
tail-aware score as
\begin{equation}
	u_i(q)=\widehat{p}_i(q)+\alpha(q)\|q_{\bar H}\|_2\|x_i\|_2 .
	\label{eq:tailscore}
\end{equation}
By Cauchy-Schwarz, the skipped contribution satisfies
\begin{equation}
	\begin{aligned}
		|q_{\bar H}^\top x_{i,\bar H}|
		\leq \|q_{\bar H}\|_2\|x_{i,\bar H}\|_2
		\leq \|q_{\bar H}\|_2\|x_i\|_2 .
	\end{aligned}
	\label{eq:tailbound}
\end{equation}
Using the full allowance for every row would be overly conservative, so ETAR
scales it by a query-dependent fraction $\alpha(q)$.

To estimate the scale of the skipped contribution, ETAR tracks the average
normalized squared value of each coordinate across active vectors. Let
$A_+=\{i\in A:\|x_i\|_2>0\}$ and $N_+=|A_+|$. For $N_+>0$, define
\begin{equation}
	\begin{aligned}
		\omega_j
		 & =\frac{1}{N_+}\sum_{i\in A_+}
		\left(\frac{x_{ij}}{\|x_i\|_2}\right)^2, \\
		\sigma_{\mathrm{tail}}^2(q)
		 & =\sum_{j\notin H} q_j^2\omega_j .
	\end{aligned}
	\label{eq:tail-statistics}
\end{equation}
Here, $\omega_j$ measures the typical fraction of a row's squared norm carried
by coordinate $j$. Weighting these fractions by the skipped query magnitudes
gives $\sigma_{\mathrm{tail}}^2(q)$, an estimate of the skipped score's second
moment under a diagonal approximation that ignores cross-coordinate
interactions. If $A_+$ is empty, we set $\omega_j=0$ for all $j$. ETAR converts
this estimate into the correction fraction
\begin{equation}
	\alpha(q)=\operatorname{clip}\!\left(
	\lambda z_{A,R}
	\frac{\sigma_{\mathrm{tail}}(q)}{\|q_{\bar H}\|_2+\epsilon},
	\alpha_{\min},\alpha_{\max}
	\right).
	\label{eq:alpha-tail}
\end{equation}
The ratio $\sigma_{\mathrm{tail}}(q)/\|q_{\bar H}\|_2$ estimates how large the
skipped contribution is relative to the query-side factor in the Cauchy-Schwarz
allowance. Multiplying this ratio by $\|x_i\|_2$ in \eqref{eq:tailscore}
therefore yields a row-scaled correction. Here, $R_A=\min\{R,|A|\}$ and
$z_{A,R}=\sqrt{2\log(|A|/R_A)}$ increases the correction according to how
selective the shortlist is. We choose $z_{A,R}$ based on a Gaussian upper-tail
approximation with tail probability $R_A/|A|$. The constant $\lambda$
calibrates the overall correction, while $\alpha_{\min}$ and
$\alpha_{\max}$ prevent it from becoming too small or too large. We set
$\epsilon=10^{-12}$ for numerical stability. The resulting $u_i(q)$ is a
heuristic candidate score, not a certified bound on the true dot product.

ETAR keeps the top-$R$ active rows by the tail-aware score and reranks them
exactly, giving
\begin{equation}
	\begin{aligned}
		S_R(q;A)
		 & =\TopOp{R}_{i\in A} u_i(q),            \\
		\widehat{I}_K(q;A)
		 & =\TopOp{K}_{i\in S_R(q;A)} q^\top x_i.
	\end{aligned}
	\label{eq:etar-output}
\end{equation}
ETAR-LM uses the same candidate set and query, but replaces the 32-bit stored
reranking row with its row-wise 8-bit reconstruction. Its full-coordinate
approximate score is
\begin{equation}
	\begin{aligned}
		\widetilde{s}_i(q)
		 & =\gamma_i\sum_{j=1}^{d} q_jc_{ij},             \\
		\widehat{I}^{\mathrm{LM}}_K(q;A)
		 & =\TopOp{K}_{i\in S_R(q;A)} \widetilde{s}_i(q).
	\end{aligned}
	\label{eq:lm-output}
\end{equation}
Thus, the two configurations differ only in the stored reranking view: ETAR
uses 32-bit vectors, while ETAR-LM uses row-major 8-bit codes and row scales.
ETAR-LM can therefore lose recall when 8-bit reranking changes the final order.
Algorithm~\ref{alg:etar-query} summarizes the query procedure.

\begin{algorithm}[H]
	\caption{ETAR query procedure}
	\label{alg:etar-query}
	\footnotesize
	\begin{algorithmic}[1]
		\Require $q$, $A$, $K$, $\rho$, $h_{\max}$, and $R\geq K$
		\Statex \textit{Query-adaptive candidate generation}
		\State Select $H$ using $\rho$ and $h_{\max}$
		\State Compute $\alpha(q)$ using \eqref{eq:alpha-tail}
		\For{$i\in A$}
		\State Compute partial score $\widehat{p}_i$ using \eqref{eq:reduced-score}
		\State Compute tail-aware score $u_i$ using \eqref{eq:tailscore}
		\EndFor
		\Statex \textit{Candidate selection and exact reranking}
		\State $S_R(q;A)\gets \TopOp{R}_{i\in A}u_i(q)$
		\State \Return $\TopOp{K}_{i\in S_R(q;A)}q^\top x_i$
	\end{algorithmic}
\end{algorithm}

\subsection{Design Rationale and Cost}

ETAR has two resource controls. The retained-energy target $\rho$ determines
the number of selected coordinates $h(q)=|H|$, up to $h_{\max}$. Smaller $\rho$
reduces candidate-generation work but may miss vectors whose scores depend on
skipped coordinates. The reranking budget $R$ controls how many candidates are
scored in the row-major view. Larger $R$ costs more but increases the chance
that the exact top-$K$ indices enter the candidate set.

For a target recall, $\rho$, $R$, and $h_{\max}$ can be calibrated on held-out
queries with exact top-$K$ labels. Higher $\rho$ reduces misses from skipped
coordinates, while larger $R$ improves candidate coverage at additional cost.
For ETAR-LM, coverage and final recall should be considered separately because
8-bit reranking may reorder candidates.

For $h(q)=|H|$, candidate arithmetic is $O(|A|h(q))$, followed by $O(Rd)$
reranking, instead of $O(|A|d)$ full dot-product arithmetic. ETAR remains
linear in $|A|$. Its benefit depends on $h(q)\ll d$ and a small fixed $R$.

Let $C$ be the allocated capacity. Excluding allocator overhead, a packed exact
table requires about $4|A|d$~bytes. ETAR requires $C(5d+13)$~bytes: $4Cd$ for
the row-major 32-bit view, $Cd$ for the column-major 8-bit scan view, and
13~bytes per row for metadata. ETAR-LM requires $C(2d+13)$~bytes for its
column-major and row-major 8-bit views. ETAR trades additional storage for less
full-precision corpus access per query, while ETAR-LM reduces that storage
overhead.

%% file: sections/03_experiments.tex
\section{Experimental Setup}
\label{sec:experiments}

We evaluate ETAR's retrieval quality and query latency, update cost under
streaming workloads, and latency and memory scaling with dataset size. Unless
otherwise noted, experiments use MIPS with $K=10$ and report online
single-query or single-operation latency under single-threaded execution.
Server experiments run on an AMD EPYC 9655 CPU. Each configuration is run five
times, with reported latencies averaged across runs and excluding setup and
evaluation overhead.

\subsection{Static Retrieval}

We use nine fixed datasets, each capped at $N=50{,}000$ stored vectors and
$Q=1{,}000$ queries. The four synthetic $d=256$ datasets are dense Gaussian,
sparse Gaussian, mixed heavy-tail, and norm-heavy. The latter three use 10\%
nonzero Gaussian entries, Student-$t_3$ entries with 35\% dropout, and Gaussian
corpus rows scaled by $\operatorname{LogNormal}(0,1.25)$, respectively. Five
datasets are from
ANN-Benchmarks~\cite{aumullerANNBenchmarksBenchmarkingTool2020}: SIFT
($d=128$), GloVe ($d=100$), LastFM ($d=65$), Fashion-MNIST ($d=784$), and
NYTimes ($d=256$). For these five, we use the first 50,000 stored vectors and
1,000 queries without additional normalization. We recompute exact top-10 MIPS
ground truth for each dataset.

We compare ETAR with exact full-precision scanning, full-coordinate 8-bit
scanning without reranking, and indexed references. The ETAR sweep uses
$\rho\in\{0.80,0.90,0.96\}$, $R\in\{50,100,200\}$, $h_{\max}=128$, row-wise
8-bit integer codes, column-major scan layout, tail-aware scoring with
$\lambda=0.75$, $\alpha_{\min}=0.05$, and $\alpha_{\max}=0.50$, and exact
reranking. ETAR-LM uses $\rho=0.90$ and $R=100$ with row-major 8-bit reranking.
The configurations differ only in whether reranking uses stored 32-bit vectors
or row-major 8-bit codes. The indexed references use fixed settings chosen to
favor recall and applied consistently across datasets: Faiss hierarchical
navigable small-world (HNSW) with $M=32$ and
$ef_{\mathrm{search}}=128$~\cite{malkovEfficientRobustApproximate2020,johnsonBillionScaleSimilaritySearch2021},
Faiss inverted-file (IVF) with $n_{\mathrm{list}}=512$ and
$n_{\mathrm{probe}}=256$~\cite{babenkoInvertedMultiIndex2015,johnsonBillionScaleSimilaritySearch2021},
Annoy with 100 trees and a search budget of
10,000~\cite{bernhardssonAnnoyApproximateNearest2013}, and ScaNN with all 1,000
leaves searched and a reorder budget of
5,000~\cite{guoAcceleratingLargeScaleInference2019}. These are fixed baseline
settings rather than recall-matched configurations.

To evaluate ETAR on a mobile device, we compare single-threaded AArch64/NEON
Android implementations of ETAR and ETAR-LM with exact scanning on a Samsung
Galaxy S25 Ultra (SM-S938B). We use the four synthetic distributions with
$N=50{,}000$ and $d=256$, with $\rho=0.90$ and $R=100$ for both ETAR variants.
After 100 warm-up queries, each method is measured five times for 10~s, with
execution order rotated across runs.

\subsection{Streaming Retrieval}

The streaming experiments interleave queries with inserts, replacements,
deletes, and maintenance events. Each workload starts with $50{,}000$ vectors
of dimension $d=256$ and runs for 5,000 simulation steps. We use five
workloads: append-only growth, query-heavy drift, balanced churn, burst ingest,
and sliding window. ETAR and ETAR-LM use $\rho=0.96$ and $R=100$. Both maintain
tail statistics over active rows and use the dynamic-table policy from
Section~\ref{sec:method}, starting with $C=1.25N$ to leave room for inserts.
Compaction is triggered after a deletion or replacement once deleted rows reach
10\% of occupied rows and at least 1,024 rows are deleted. The first three
workloads use query/insert/replacement/delete mixes of 85/15/0/0 (append-only),
95/3/1/1 (query-heavy), and 70/10/10/10 (balanced). Burst ingest uses 92/4/2/2,
switching to 20/70/5/5 for 100 steps every 500 steps. Sliding window cycles
through insert, delete-oldest, and query. Recall is measured every 250 updates
on 25 fixed queries. Append-only growth, query-heavy drift, and sliding window
use $\mathcal{N}(0,\mathbf{I})$ initial rows and $\mathcal{N}(0.5,\mathbf{I})$
streamed rows and queries. Balanced churn multiplies Gaussian rows by
$\operatorname{LogNormal}(0,1)$ scales. Burst ingest uses
$\mathcal{N}(0,0.25^2\mathbf{I})$ rows with one coordinate offset by
$\mathcal{N}(6,1)$.

The dynamic Faiss references support native insertion, while deletes use
external live-set filtering until rebuild and replacements delete the old item
before inserting the new one. We rebuild these indexes every 500 updates. We
additionally include HNSWlib as a dynamic MIPS baseline with native update
support, using $M=32$, $ef_{\mathrm{construction}}=200$, and
$ef_{\mathrm{search}}=512$, with marked deletions and no scheduled rebuilds.

We also include a high-churn stress experiment to measure ETAR's compaction and
table-maintenance costs directly. It uses the same initial size and dimension,
but runs for 20,000 steps with a delete- and replacement-heavy mix: 40\% query,
15\% insert, 15\% replacement, and 30\% delete. ETAR uses the same query
configuration as in the main streaming experiment. The baselines retain the
update policies above.

\subsection{Memory and Scaling Profiles}

Analytical representation size includes stored vector views, compact scan
codes, row scales, norms, deletion markers, row identifiers, and allocated
capacity. We also record process resident-set size (RSS) immediately before and
after method construction. The RSS build delta is an implementation-level
profile rather than an exact device-memory requirement because the runtime,
libraries, allocator, loaded data, and temporary arrays also affect the process
footprint.

For the scaling profile, we use dense Gaussian data with $d=256$, $Q=50$, and
$N\in\{50{,}000,100{,}000,250{,}000,500{,}000,1{,}000{,}000\}$. Each
method-size pair is measured in a fresh process. ETAR and ETAR-LM use $C=N$,
$\rho=0.96$, $R=200$, and $h_{\max}=192$. The profile also includes exact
scanning, Faiss HNSW, and Faiss IVF. The configurations are not recall-matched.

\subsection{Metrics}

For a query set $\mathcal{Q}$, let $A(q)$ be the active set when query $q$ is
issued, and let $K_q=\min\{K,|A(q)|\}$. The exact target is $I_K(q;A(q))$ from
\eqref{eq:mips}, and the returned set is $\widehat{I}_K(q;A(q))$. We report
\begin{equation}
	\mathrm{Recall@}K
	=
	\frac{1}{|\mathcal{Q}|}
	\sum_{q\in\mathcal{Q}}
	\frac{|I_K(q;A(q))\cap \widehat{I}_K(q;A(q))|}{K_q} .
	\label{eq:recall}
\end{equation}

For diagnostic analysis, candidate coverage is defined using \eqref{eq:recall}
with $\widehat{I}_K(q;A(q))$ replaced by $S_R(q;A(q))$. For ETAR-LM, the gap
between coverage and recall reflects reordering from 8-bit quantization during
final scoring. We report coverage where this distinction is informative. Event
time includes queries, updates, and maintenance. Initial build time measures
method construction from the initial 50,000 rows and is excluded from event
time. We also report query latency, maintenance latency, analytical
representation size, and RSS build delta. In all tables, bold marks the best
directly comparable value within each dataset, workload, or column.

%% file: sections/04_results.tex
\section{Results and Discussion}
\label{sec:results}

\subsection{Static Retrieval}

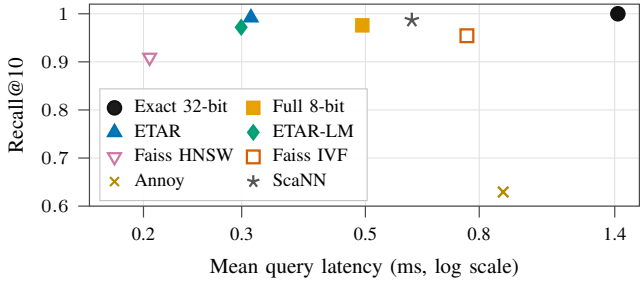
\begin{figure}[t]
	\centering
	\input{figures/static_recall_latency.tex}
	\caption{Mean recall-latency tradeoff across nine static datasets.}
	\label{fig:static-recall-latency}
\end{figure}

Fig.~\ref{fig:static-recall-latency} summarizes the recall-latency tradeoff
averaged across the nine static datasets. Exact full-precision scanning reaches
perfect Recall@10 at about 1.42~ms/query. At the representative setting
$\rho=0.90$ and $R=100$, ETAR reaches 99.2\% Recall@10 at about 0.31~ms/query,
a 4.5$\times$ speedup over exact scanning. A more conservative setting,
$\rho=0.96$ and $R=200$, reaches 99.97\% Recall@10 at about 0.37~ms/query.
ETAR-LM reaches 97.2\% Recall@10 at about 0.30~ms/query while using a smaller
representation. Faiss HNSW is faster on average under this protocol but
achieves lower Recall@10, whereas ScaNN reaches high recall at higher latency
than these ETAR settings.

Results on the mobile device show a similar trend: ETAR and ETAR-LM achieve
mean Recall@10 of 97.59--100\% and 96.37--99.42\%, respectively, while
providing 1.76--6.88$\times$ and 1.86--7.14$\times$ speedups over exact
scanning, with the largest gains on sparse Gaussian.

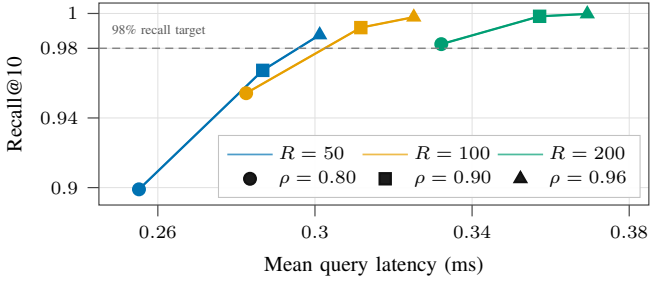
\begin{figure}[t]
	\centering
	\input{figures/etar_sweep.tex}
	\caption{Mean ETAR recall-latency sweep across nine datasets by $R$ and $\rho$.}
	\label{fig:etar-sweep}
\end{figure}

Fig.~\ref{fig:etar-sweep} shows the ETAR sweep in recall-latency space. Along
each curve, $\rho$ increases from 0.80 to 0.90 to 0.96. At $R=100$, Recall@10
rises from 95.42\% to 99.19\% and 99.79\%, while latency rises from 0.282 to
0.312 and 0.325~ms/query. Lower $\rho$ therefore increases the chance of
skipping important score contributions, while a larger $R$ can recover more
candidates at additional reranking cost.

Results at the common $\rho=0.90$, $R=100$ setting further show that recall
depends on how query magnitude is distributed across coordinates, not on
dimension alone. Dense Gaussian data reaches 97.19\% Recall@10 while selecting
114.4 coordinates on average, whereas 10\%-sparse data selects only 12.3
coordinates and reaches 99.68\%. At fixed $\rho$, queries concentrated in fewer
coordinates can therefore reach the target with fewer selected coordinates than
more diffuse queries.

Table~\ref{tab:target-recall} provides a per-dataset comparison at high recall.
For each dataset, it reports the fastest ETAR sweep point with Recall@10 $\geq
	0.98$ and applies the same threshold to the four fixed indexed baselines. ETAR
is faster than exact scanning on all nine datasets, with speedups up to
9.83$\times$. The gains are smaller on lower-dimensional datasets such as
LastFM and GloVe, where full dot products involve fewer coordinates. The
selected indexed baseline is faster than ETAR on three datasets and slower on
the remaining six. Because these baselines use fixed rather than per-dataset
tuning, the comparison should not be interpreted as a universal ranking.

\input{tables/table_target_recall.tex}

Table~\ref{tab:static-ablation} summarizes the main design ablations. Removing
the tail term lowers average Recall@10 from 99.19\% to 98.17\%, with the
largest benefit on mixed heavy-tail data, where recall rises from 94.17\% to
99.98\%. This suggests that the tail correction is most useful when the skipped
score contribution can be predicted from row norm and coordinate-wise second
moment. A fixed-tail variant with $\alpha=0.10$ nearly matches the derived
correction on average while running slightly faster. This suggests that the
tail allowance accounts for most of the average recall gain, with only a small
difference between the two variants in this comparison. Removing reranking
causes the largest recall drop, confirming that the reduced scan should select
candidates rather than determine the final order.

With deterministic tie-breaking and exact reranking, ETAR's 99.19\% candidate
coverage matches its final recall. ETAR-LM uses the same candidate sets but
reaches 97.19\% final recall. The gap is due to quantization during 8-bit
reranking.

\input{tables/table_static_ablation.tex}

\subsection{Streaming Retrieval}

Table~\ref{tab:dynamic-operations} highlights the main dynamic tradeoff. ETAR
and ETAR-LM build in under 0.05~s and incur at most 0.007~ms per update. ETAR
averages 0.523~ms/query, while Faiss HNSW queries faster but has more expensive
updates and 21.3~s average rebuilds. HNSWlib avoids scheduled rebuilds but has
higher query and update latency than ETAR, averaging 1.79~ms/query and at least
0.935~ms per insert or replacement. Faiss IVF also requires periodic rebuilds
that include retraining.

\input{tables/table_dynamic_operations.tex}

Table~\ref{tab:dynamic-workloads} shows that ETAR maintains 100\% Recall@10 at
all checkpoints, while ETAR-LM averages 99.1\%. Faiss HNSW generally has the
lowest query latency, but periodic rebuilds increase total event time under
heavier update rates. HNSWlib avoids scheduled rebuilds and maintains at least
99.6\% recall, but at higher query latency. Overall, ETAR or ETAR-LM achieves
the lowest event time in four of the five workloads.

\input{tables/table_dynamic_workloads.tex}

Faiss rebuilds introduce large latency spikes, whereas HNSWlib requires no
scheduled rebuilds. Neither ETAR configuration triggers compaction in these
workloads.

\subsection{Compaction Stress Test}

The 20,000-step high-churn experiment uses 30\% deletes and 15\% replacements
to stress ETAR's compaction mechanism. Table~\ref{tab:compaction-stress} shows
that ETAR and ETAR-LM each compact only once under high churn, taking 31.4 and
19.2~ms, respectively. HNSWlib avoids scheduled maintenance and reaches 100\%
Recall@10, but its total event time is 16.25~s compared with 4.24~s for ETAR.
In contrast, Faiss HNSW and IVF each rebuild 24 times, leading to much larger
total event times.

\input{tables/table_compaction_stress.tex}

\subsection{Memory and Scaling}

Table~\ref{tab:memory-profile} reports analytical representation size for
$d=256$ and $C=N$. At $N=50{,}000$, ETAR uses 61.7~MiB, or 1.26$\times$ exact,
while ETAR-LM uses 25.0~MiB, or 0.51$\times$ exact. ETAR primarily reduces
query computation rather than storage, whereas ETAR-LM also reduces storage at
the cost of a moderate recall loss.

\input{tables/table_memory_profile.tex}

\begin{figure}[t]
	\centering
	\input{figures/scaling_profile.tex}
	\caption{Scaling profile ($d=256$, not recall-matched, log axes).}
	\label{fig:scaling-profile}
\end{figure}
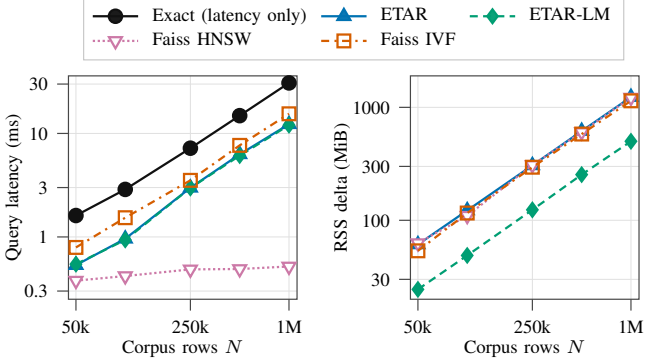

Fig.~\ref{fig:scaling-profile} shows approximately linear query-latency growth
with $N$ for exact scanning and both ETAR configurations. At one million rows,
exact scanning takes 30.8~ms/query, compared with 12.3~ms for ETAR at 100\%
Recall@10 and 12.0~ms for ETAR-LM at 97.4\%. The fixed HNSW and IVF settings
achieve lower recall at this scale. RSS build delta also grows roughly
linearly, with ETAR-LM remaining the smallest at 500.6~MiB, consistent with the
analytical representation sizes in Table~\ref{tab:memory-profile}.

%% file: figures/static_recall_latency.tex
\begin{tikzpicture}
	\begin{axis}[
			paper axis,
			width=\linewidth,
			height=0.48\linewidth,
			xmode=log,
			xmin=0.16,
			xmax=1.55,
			ymin=0.60,
			ymax=1.02,
			xtick pos=bottom,
			ytick pos=left,
			xtick={0.2,0.3,0.5,0.8,1.4},
			xticklabels={0.2,0.3,0.5,0.8,1.4},
			ytick={0.6,0.7,0.8,0.9,1.0},
			xlabel={Mean query latency (ms, log scale)},
			ylabel={Recall@10},
			legend columns=2,
			legend style={at={(0.018,0.025)},anchor=south west,column sep=3pt},
		]
		\addplot[
			scatter,
			only marks,
			scatter src=explicit symbolic,
			scatter/classes={
					exact={mark=*,draw=etarBlack,fill=etarBlack,mark options={solid},mark size=2.5pt},
					full8={mark=square*,draw=etarOrange,fill=etarOrange,mark options={solid},mark size=2.3pt},
					etar={mark=triangle*,draw=etarBlue,fill=etarBlue,mark options={solid},mark size=2.8pt},
					etarlm={mark=diamond*,draw=etarTeal,fill=etarTeal,mark options={solid},mark size=2.6pt},
					hnsw={mark=triangle*,draw=etarPurple,fill=white,mark options={solid},mark size=2.8pt,rotate=180},
					ivf={mark=square,draw=etarSky,fill=white,mark options={solid},mark size=2.4pt},
					annoy={mark=x,draw=etarGold,mark options={solid},mark size=2.8pt,line width=0.9pt},
					scann={mark=star,draw=etarGray,fill=etarGray,mark options={solid},mark size=2.8pt}
				},
			forget plot
		] table[x=latency_ms,y=recall_at_k,meta=method_key,col sep=comma]
			{figures/data/static_summary.csv};

		\addlegendimage{only marks,/tikz/exact mark}
		\addlegendentry{Exact 32-bit}
		\addlegendimage{only marks,/tikz/full8 mark}
		\addlegendentry{Full 8-bit}
		\addlegendimage{only marks,/tikz/etar mark}
		\addlegendentry{ETAR}
		\addlegendimage{only marks,/tikz/etarlm mark}
		\addlegendentry{ETAR-LM}
		\addlegendimage{only marks,/tikz/hnsw mark}
		\addlegendentry{Faiss HNSW}
		\addlegendimage{only marks,/tikz/ivf mark}
		\addlegendentry{Faiss IVF}
		\addlegendimage{only marks,mark=x,mark options={solid,draw=etarGold},mark size=2.4pt,line width=0.8pt}
		\addlegendentry{Annoy}
		\addlegendimage{only marks,mark=star,mark options={solid,draw=etarGray,fill=etarGray},mark size=2.4pt}
		\addlegendentry{ScaNN}
	\end{axis}
\end{tikzpicture}

%% file: figures/etar_sweep.tex
\begin{tikzpicture}
	\begin{axis}[
			paper axis,
			width=\linewidth,
			height=0.48\linewidth,
			xmin=0.245,
			xmax=0.385,
			ymin=0.890,
			ymax=1.006,
			xtick pos=bottom,
			ytick pos=left,
			xtick={0.26,0.30,0.34,0.38},
			ytick={0.90,0.94,0.98,1.00},
			xlabel={Mean query latency (ms)},
			ylabel={Recall@10},
			legend columns=3,
			legend style={at={(0.985,0.035)},anchor=south east},
		]
		\addplot[draw=etarBlue!70,line width=0.65pt,no marks,forget plot]
		table[x=latency_ms,y=recall_at_k,col sep=comma,
				restrict expr to domain={\thisrow{R}}{49:51}]
			{figures/data/etar_sweep.csv};
		\addplot[
			draw=etarBlue,
			scatter,
			scatter src=explicit symbolic,
			scatter/classes={
					0.800000000={mark=*,draw=etarBlue,fill=etarBlue,mark options={solid},mark size=2.2pt},
					0.900000000={mark=square*,draw=etarBlue,fill=etarBlue,mark options={solid},mark size=2.1pt},
					0.960000000={mark=triangle*,draw=etarBlue,fill=etarBlue,mark options={solid},mark size=2.5pt}
				},
			forget plot
		]
		table[x=latency_ms,y=recall_at_k,col sep=comma,
				meta=rho,
				restrict expr to domain={\thisrow{R}}{49:51}]
			{figures/data/etar_sweep.csv};

		\addplot[draw=etarOrange!70,line width=0.65pt,no marks,forget plot]
		table[x=latency_ms,y=recall_at_k,col sep=comma,
				restrict expr to domain={\thisrow{R}}{99:101}]
			{figures/data/etar_sweep.csv};
		\addplot[
			draw=etarOrange,
			scatter,
			scatter src=explicit symbolic,
			scatter/classes={
					0.800000000={mark=*,draw=etarOrange,fill=etarOrange,mark options={solid},mark size=2.2pt},
					0.900000000={mark=square*,draw=etarOrange,fill=etarOrange,mark options={solid},mark size=2.1pt},
					0.960000000={mark=triangle*,draw=etarOrange,fill=etarOrange,mark options={solid},mark size=2.5pt}
				},
			forget plot
		]
		table[x=latency_ms,y=recall_at_k,col sep=comma,
				meta=rho,
				restrict expr to domain={\thisrow{R}}{99:101}]
			{figures/data/etar_sweep.csv};

		\addplot[draw=etarTeal!70,line width=0.65pt,no marks,forget plot]
		table[x=latency_ms,y=recall_at_k,col sep=comma,
				restrict expr to domain={\thisrow{R}}{199:201}]
			{figures/data/etar_sweep.csv};
		\addplot[
			draw=etarTeal,
			scatter,
			scatter src=explicit symbolic,
			scatter/classes={
					0.800000000={mark=*,draw=etarTeal,fill=etarTeal,mark options={solid},mark size=2.2pt},
					0.900000000={mark=square*,draw=etarTeal,fill=etarTeal,mark options={solid},mark size=2.1pt},
					0.960000000={mark=triangle*,draw=etarTeal,fill=etarTeal,mark options={solid},mark size=2.5pt}
				},
			forget plot
		]
		table[x=latency_ms,y=recall_at_k,col sep=comma,
				meta=rho,
				restrict expr to domain={\thisrow{R}}{199:201}]
			{figures/data/etar_sweep.csv};

		\addlegendimage{draw=etarBlue!70,line width=0.65pt}
		\addlegendentry{$R=50$}
		\addlegendimage{draw=etarOrange!70,line width=0.65pt}
		\addlegendentry{$R=100$}
		\addlegendimage{draw=etarTeal!70,line width=0.65pt}
		\addlegendentry{$R=200$}
		\addlegendimage{only marks,mark=*,mark options={solid,draw=etarBlack,fill=etarBlack},mark size=2.2pt}
		\addlegendentry{$\rho=0.80$}
		\addlegendimage{only marks,mark=square*,mark options={solid,draw=etarBlack,fill=etarBlack},mark size=2.1pt}
		\addlegendentry{$\rho=0.90$}
		\addlegendimage{only marks,mark=triangle*,mark options={solid,draw=etarBlack,fill=etarBlack},mark size=2.5pt}
		\addlegendentry{$\rho=0.96$}

		\addplot[draw=black!48,densely dashed,line width=0.55pt,forget plot]
		coordinates {(0.245,0.98) (0.385,0.98)};
		\node[font=\tiny,anchor=south west,text=black!65]
		at (axis cs:0.246,0.9812) {98\% recall target};
	\end{axis}
\end{tikzpicture}

%% file: tables/table_target_recall.tex
\begin{table*}[t]
	\centering
	\caption{Per-dataset ETAR and indexed settings at Recall@10 $\geq0.98$.}
	\label{tab:target-recall}
	\setlength{\tabcolsep}{3.4pt}
	\begin{tabular*}{\textwidth}{@{\extracolsep{\fill}}l r r r r r r l r r}
		\toprule
		& & \multicolumn{5}{c}{ETAR} & \multicolumn{3}{c}{Indexed reference} \\
		\cmidrule(lr){3-7}\cmidrule(lr){8-10}
		Dataset & $d$ & $\rho$ & $R$ & Recall@10 & Query (ms) $\downarrow$ & Speedup vs. exact & Method & Recall@10 & Query (ms) $\downarrow$ \\
		\midrule
		Dense Gaussian & 256 & .96 & 100 & .991 & \textbf{.373} & $3.65\times$ & ScaNN & 1.000 & .621 \\
		Sparse Gaussian & 256 & .90 & 50 & .987 & \textbf{.185} & $7.31\times$ & ScaNN & 1.000 & .604 \\
		Mixed heavy-tail & 256 & .80 & 50 & .998 & \textbf{.210} & $6.23\times$ & Faiss IVF & .980 & .701 \\
		Norm-heavy & 256 & .80 & 50 & .991 & .280 & $4.54\times$ & Faiss HNSW & .997 & \textbf{.209} \\
		SIFT & 128 & .96 & 50 & .991 & .210 & $1.95\times$ & Faiss HNSW & 1.000 & \textbf{.081} \\
		GloVe & 100 & .96 & 50 & .997 & \textbf{.241} & $1.36\times$ & ScaNN & 1.000 & .277 \\
		LastFM & 65 & .90 & 50 & .991 & .199 & $1.37\times$ & Faiss HNSW & .998 & \textbf{.048} \\
		Fashion-MNIST & 784 & .80 & 100 & .992 & \textbf{.518} & $\textbf{9.83}\times$ & ScaNN & 1.000 & 1.605 \\
		NYTimes & 256 & .90 & 100 & .995 & \textbf{.352} & $3.88\times$ & ScaNN & .999 & .689 \\
		\bottomrule
	\end{tabular*}
\end{table*}

%% file: tables/table_static_ablation.tex
\begin{table}[t]
	\centering
	\caption{Mean ablation results across nine datasets at $\rho=0.90$, with $R=100$ for reranking.}
	\label{tab:static-ablation}
	\setlength{\tabcolsep}{3.2pt}
	\begin{tabular*}{\columnwidth}{@{\extracolsep{\fill}}lrrr}
		\toprule
		Variant & Recall@10 $\uparrow$ & Coverage $\uparrow$ & Query (ms) $\downarrow$ \\
		\midrule
		ETAR & \textbf{.9919} & \textbf{.9919} & .312 \\
		No tail term & .9817 & .9817 & .344 \\
		Fixed tail & .9915 & .9915 & .287 \\
		No reranking & .6545 & .6545 & \textbf{.246} \\
		ETAR-LM & .9719 & \textbf{.9919} & .299 \\
		\bottomrule
	\end{tabular*}
\end{table}

%% file: tables/table_dynamic_operations.tex
\begin{table}[t]
	\centering
	\caption{Streaming build, operation, and maintenance costs.}
	\label{tab:dynamic-operations}
	\setlength{\tabcolsep}{1.6pt}
	\begin{tabular*}{\columnwidth}{@{\extracolsep{\fill}}lrrrrrl}
		\toprule
		Method & \shortstack{Build\\(s) $\downarrow$} & \shortstack{Query\\(ms) $\downarrow$} & \shortstack{Ins.\\(ms) $\downarrow$} &
		\shortstack{Repl.\\(ms) $\downarrow$} & \shortstack{Del.\\(ms) $\downarrow$} & Maintenance \\
		\midrule
		ETAR & .047 & .523 & \textbf{.005} & .007 & .002 & \textbf{0 events} \\
		ETAR-LM & \textbf{.041} & .516 & \textbf{.005} & \textbf{.006} & .002 & \textbf{0 events} \\
		Faiss HNSW & 19.9 & \textbf{.255} & .640 & .674 & \textbf{$<.001$} & 21.3~s rebuild \\
		HNSWlib & 47.6 & 1.792 & .935 & 1.162 & .005 & \textbf{0 events} \\
		Faiss IVF & 3.26 & 1.163 & .020 & .023 & .001 & 3.26~s rebuild \\
		\bottomrule
	\end{tabular*}
\end{table}

%% file: tables/table_dynamic_workloads.tex
\begin{table}[t]
	\centering
	\caption{Streaming retrieval results by workload.}
	\label{tab:dynamic-workloads}
	\setlength{\tabcolsep}{2.2pt}
	\renewcommand{\arraystretch}{0.92}
	\begin{tabular*}{\columnwidth}{@{\extracolsep{\fill}}llrrr}
		\toprule
		Workload & Method & Recall@10 $\uparrow$ & Query (ms) $\downarrow$ & Event (s) $\downarrow$ \\
		\midrule
		Append-only & ETAR & \textbf{1.000} & .469 & \textbf{2.00} \\
		& ETAR-LM & .991 & .484 & 2.06 \\
		& Faiss HNSW & \textbf{1.000} & \textbf{.186} & 16.87 \\
		& HNSWlib & \textbf{1.000} & 1.688 & 7.74 \\
		& Faiss IVF & .929 & 1.113 & 8.06 \\
		\addlinespace[1.5pt]
		Query-heavy drift & ETAR & \textbf{1.000} & .547 & 2.60 \\
		& ETAR-LM & .992 & .538 & 2.56 \\
		& Faiss HNSW & .984 & \textbf{.324} & \textbf{1.62} \\
		& HNSWlib & \textbf{1.000} & 2.782 & 13.53 \\
		& Faiss IVF & .804 & 1.146 & 5.45 \\
		\addlinespace[1.5pt]
		Balanced churn & ETAR & \textbf{1.000} & .537 & 1.89 \\
		& ETAR-LM & .991 & .533 & \textbf{1.87} \\
		& Faiss HNSW & .997 & \textbf{.372} & 154.97 \\
		& HNSWlib & .996 & 1.277 & 5.35 \\
		& Faiss IVF & .953 & 1.221 & 13.80 \\
		\addlinespace[1.5pt]
		Burst ingest & ETAR & \textbf{1.000} & .481 & 1.90 \\
		& ETAR-LM & .992 & .478 & \textbf{1.89} \\
		& Faiss HNSW & .987 & \textbf{.153} & 7.22 \\
		& HNSWlib & \textbf{1.000} & 2.266 & 9.88 \\
		& Faiss IVF & \textbf{1.000} & 1.181 & 11.29 \\
		\addlinespace[1.5pt]
		Sliding window & ETAR & \textbf{1.000} & .582 & .98 \\
		& ETAR-LM & .991 & .548 & \textbf{.92} \\
		& Faiss HNSW & .998 & \textbf{.239} & 95.69 \\
		& HNSWlib & \textbf{1.000} & .946 & 2.33 \\
		& Faiss IVF & .983 & 1.154 & 21.47 \\
		\bottomrule
	\end{tabular*}
\end{table}

%% file: tables/table_compaction_stress.tex
\begin{table}[t]
	\centering
	\caption{High-churn compaction stress results.}
	\label{tab:compaction-stress}
	\setlength{\tabcolsep}{1.0pt}
	\begin{tabular*}{\columnwidth}{@{\extracolsep{\fill}}lrrrrr}
		\toprule
		Method & \shortstack{Build\\(s) $\downarrow$} &
		\shortstack{Query\\(ms) $\downarrow$} & \shortstack{Maintenance\\count$\times$mean $\downarrow$} &
		\shortstack{Event\\(s) $\downarrow$} & \shortstack{Recall@10 $\uparrow$} \\
		\midrule
		ETAR & \textbf{.042} & .517 & 1$\times$31.4~ms & 4.24 & \textbf{1.0000} \\
		ETAR-LM & .046 & .509 & 1$\times$19.2~ms & \textbf{4.15} & .9915 \\
		Faiss HNSW & 50.1 & \textbf{.403} & 24$\times$49.3~s & 1195.3 & .9982 \\
		HNSWlib & 26.2 & 1.358 & \textbf{0 events} & 16.25 & \textbf{1.0000} \\
		Faiss IVF & 3.12 & 1.140 & 24$\times$2.97~s & 80.6 & .9593 \\
		\bottomrule
	\end{tabular*}
\end{table}

%% file: tables/table_memory_profile.tex
\begin{table}[t]
	\centering
	\caption{Analytical representation size in MiB for $d=256$ and $C=N$.}
	\label{tab:memory-profile}
	\setlength{\tabcolsep}{3.0pt}
	\begin{tabular*}{\columnwidth}{@{\extracolsep{\fill}}l l c r r}
		\toprule
		Mode & Stored vector views & Bytes $\downarrow$ & 50k $\downarrow$ & 1M $\downarrow$ \\
		\midrule
		Exact & row 32-bit & $4Nd$ & 48.8 & 976.6 \\
		ETAR & col 8-bit + row 32-bit & $N(5d+13)$ & 61.7 & 1233.1 \\
		ETAR-LM & col 8-bit + row 8-bit & \textbf{$N(2d+13)$} & \textbf{25.0} & \textbf{500.7} \\
		\bottomrule
	\end{tabular*}
\end{table}

%% file: figures/scaling_profile.tex
\begin{tikzpicture}
	\begin{groupplot}[
			group style={
					group size=2 by 1,
					horizontal sep=1.5cm,
				},
			paper axis,
			width=0.52\columnwidth,
			height=0.25\textwidth,
			xmode=log,
			ymode=log,
			xtick pos=bottom,
			ytick pos=left,
			xmin=45000,
			xmax=1120000,
			xtick={50000,250000,1000000},
			xticklabels={50k,250k,1M},
			xlabel style={font=\scriptsize,yshift=4pt},
			xlabel={Corpus rows $N$},
		]

		\nextgroupplot[
			ymin=0.25,
			ymax=36,
			ytick={0.3,1,3,10,30},
			yticklabels={0.3,1,3,10,30},
			ylabel near ticks,
			ylabel style={font=\scriptsize,yshift=-6pt},
			ylabel={Query latency (ms)},
			legend columns=3,
			legend style={
					column sep=3pt,
				},
			legend to name=scalinglegend,
		]
		\addplot[exact line] table[x=N,y=exact_latency_ms,col sep=comma] {figures/data/scaling_profile.csv};
		\addlegendentry{Exact (latency only)}
		\addplot[etar line] table[x=N,y=etar_latency_ms,col sep=comma] {figures/data/scaling_profile.csv};
		\addlegendentry{ETAR}
		\addplot[etarlm line] table[x=N,y=etarlm_latency_ms,col sep=comma] {figures/data/scaling_profile.csv};
		\addlegendentry{ETAR-LM}
		\addplot[hnsw line] table[x=N,y=hnsw_latency_ms,col sep=comma] {figures/data/scaling_profile.csv};
		\addlegendentry{Faiss HNSW}
		\addplot[ivf line] table[x=N,y=ivf_latency_ms,col sep=comma] {figures/data/scaling_profile.csv};
		\addlegendentry{Faiss IVF}

		\nextgroupplot[
			ymin=20,
			ymax=1900,
			ytick={30,100,300,1000},
			yticklabels={30,100,300,1000},
			ylabel near ticks,
			ylabel style={font=\scriptsize,yshift=-6pt},
			ylabel={RSS delta (MiB)},
		]
		\addplot[etar line] table[x=N,y=etar_rss_delta_mib,col sep=comma] {figures/data/scaling_profile.csv};
		\addplot[etarlm line] table[x=N,y=etarlm_rss_delta_mib,col sep=comma] {figures/data/scaling_profile.csv};
		\addplot[hnsw line] table[x=N,y=hnsw_rss_delta_mib,col sep=comma] {figures/data/scaling_profile.csv};
		\addplot[ivf line] table[x=N,y=ivf_rss_delta_mib,col sep=comma] {figures/data/scaling_profile.csv};
	\end{groupplot}
	\node[anchor=south] at ([yshift=0.08cm]$(group c1r1.north)!0.5!(group c2r1.north)$)
	{\pgfplotslegendfromname{scalinglegend}};
\end{tikzpicture}

%% file: sections/05_limitations.tex
\section{Limitations and Future Work}
\label{sec:limitations}

ETAR scans all active rows, so query latency grows linearly with $N$. Its
heuristic score does not guarantee that every true top-$K$ row enters $S_R$,
while ETAR-LM may reorder candidates during 8-bit reranking. Resizing and
compaction can cause maintenance spikes, and parameter settings do not adapt to
workload changes. The evaluation uses fixed indexed-baseline settings and a
single periodic rebuild policy, and does not include other scan-based MIPS
methods. Edge-device validation is limited to one ARM-based mobile device and
static synthetic workloads. Future work includes larger corpora, batched
queries, dynamic edge workloads, energy measurements on edge and embedded
systems, and adaptive selection of $\rho$, $R$, and maintenance policies.

%% file: sections/06_conclusion.tex
\section{Conclusion}
\label{sec:conclusion}

ETAR shows that dynamic MIPS can reduce full-scan query work without
auxiliary-index maintenance. By focusing on the highest-energy query
coordinates and reranking a bounded set of candidates, it preserves simple
updates while achieving high recall and substantial speedups across server and
mobile experiments. ETAR also maintains 100\% Recall@10 across the streaming
workloads without index rebuilds, while ETAR-LM extends the design to
lower-storage retrieval. Overall, ETAR offers a practical middle ground between
full-vector scanning and index-based retrieval for dynamic MIPS workloads.